\documentclass[10pt,final,twocolumn,showpacs,preprintnumbers,a4paper,amsmath,amssymb,rmp]{revtex4}

\usepackage[T1]{fontenc}
\usepackage{times,mathptmx,eucal}
\usepackage{bm}
\usepackage{hyperref}

\usepackage{ntheorem}
 \theoremstyle{plain}
 {\theorembodyfont{\rmfamily}
  \theoremseparator{.}
  
  \newtheorem{thm}{Theorem}[section]
  \theoremstyle{plain}
  
  \theoremstyle{plain} 
  \theoremstyle{plain}

  \theoremstyle{plain}
  \newtheorem{rem}{Remark}[section]}

\renewcommand*{\vec}[1]{\mathbf{#1}}
\usepackage{xspace}
\newcommand*{\R}{\ensuremath{\mathbb{R}}\xspace}
\newcommand*{\cross}{\bm{\times}}
\newcommand*{\grad}{\bm{\nabla}}
\newcommand*{\curl}{\grad\cross}
\renewcommand*{\div}{\grad\cdot}
\renewcommand*{\d}{\mathrm{d}}
\newcommand*{\dt}{\mathrm{d}t}
\newcommand*{\dx}{\mathrm{d}x}
\newcommand*{\ddt}[1]{\frac{\d{#1}}{\dt}}
\newcommand*{\pdds}[1]{\frac{\partial{#1}}{\partial s}}
\newcommand*{\pddt}[1]{\frac{\partial{#1}}{\partial t}}
\newcommand*{\pddx}[1]{\frac{\partial{#1}}{\partial x}}
\newcommand*{\pddy}[1]{\frac{\partial{#1}}{\partial y}}
\newcommand*{\pddz}[1]{\frac{\partial{#1}}{\partial z}}

\begin{document}

\preprint{Submitted to J. Math.\ Phys. March 5, 2007}

\title{Conservation laws for the Maxwell-Dirac equations with
a dual Ohm's law}

\author{Nail H.~Ibragimov}
 \email{nib@bth.se}
\author{Raisa Khamitova}
 \affiliation{Department of Mathematics and Science,
 Research Centre ALGA: Advances in Lie Group Analysis,
 Blekinge Institute of Technology,
 SE-371\,79 Karlskrona, Sweden}

\author{Bo Thid\'{e}}
 \email{bt@irfu.se}
 \affiliation{Swedish Institute of Space Physics,
 \r{A}ngstr\"om Laboratory,  P.\,O.~Box~537,
 SE-751\,21 Uppsala, Sweden}
 \altaffiliation{Also at LOIS Space Centre,
 School of Mathematics and Systems Engineering, 
 V\"{a}xj\"{o} University, SE-351\,95 V\"axj\"o, Sweden}

\begin{abstract}
Using a general theorem on conservation laws for arbitrary differential
equations proved by Ibragimov, we have derived conservation laws for
Dirac's symmetrized Maxwell-Lorentz equations under the assumption that
both the electric and magnetic charges obey linear conductivity laws
(dual Ohm's law).  We find that this linear system allows for
conservation laws which are non-local in time.
\end{abstract}

\maketitle

\section{Introduction}

In all areas of physics, conservation laws are essential since
they allow us to draw conclusions of a physical system under study
in an efficient way.

Electrodynamics, in terms of the standard Maxwell electromagnetic
equations for fields in vacuum, exhibit a rich set of symmetries to
which conserved quantities are associated.  Recently, there has been a
renewed interest in the utilisation of such quantities.  Here we use a
theorem of \textcite{Ibragimov:JMAA:2006b} to derive conservation laws
for Dirac's symmetric version of the Maxwell-Lorentz microscopic
equations, allowing for magnetic charges and magnetic currents, where
the latter, just as electric currents, are assumed to be described by a
linear relationship between the field and the current, i.e., an Ohm's
law.  The method of \textcite{Ibragimov:JMAA:2006b} produces two new
adjoint vector fields which fulfil Maxwell-like equations.  In
particular, we obtain conservation laws for the symmetrized
electromagnetic field which are non-local in time.

\section{Preliminaries}

\subsection{Notation}

We will use the following notation (see, e.g., Ref.\
\onlinecite{Ibragimov:1999}).  Let $x=(x^1,\ldots,x^n)$ be
independent variables and $u=(u^1,\ldots,u^m)$ be dependent variables.
The set of the first-order partial derivatives $u^{\alpha}_i=\partial
u^\alpha/\partial x^i$ will be denoted by $u_{(1)}=\{u^{\alpha}_i\}$,
where $\alpha=1,\ldots.,m$ and $i,j,\ldots=1,\ldots,n$.  The symbol
$D_i$ denotes the total differentiation with respect to the variable
$x^i$:
\begin{equation*}
 D_i = \frac{\partial}{\partial x^i} +
 u^\alpha_{i}\frac{\partial}{\partial u^\alpha} +
 u^\alpha_{ij}\frac{\partial}{\partial u^\alpha_j} +
 u^\alpha_{ijk}\frac{\partial}{\partial u^\alpha_{jk}} + \cdots
\end{equation*}
We employ the usual convention of summation in repeated
indices.

Recall that a necessary condition for extrema of a variational integral
\begin{equation}
 \label{F1}
 \int_V\mathcal{L}(x,u,u_{(1)})\,\dx
\end{equation}
with a Lagrangian $\mathcal{L}(x,u,u_{(1)})$, depending on
first-order derivatives, is given by the Euler-Lagrange equations
\begin{equation}
 \frac{\delta\mathcal{L}}{\delta u^\alpha} \equiv
 \frac{\partial{\mathcal{L}}}{\partial u^\alpha}
 - D_i\bigg(\frac{\partial{\mathcal{L}}}{\partial u^\alpha_i}\bigg)
 = 0, \quad \alpha = 1,\ldots, m.
 \label{F2}
\end{equation}

We will understand by a \emph{symmetry} of a certain system of
differential equations a generator
\begin{equation}
\label{F3}
 X=\xi^i(x,u)\frac{\partial}{\partial x^i} + \eta^\alpha (x,u)
 \frac{\partial}{\partial u^\alpha}
\end{equation}
of a continuous transformation group admitted by differential equations
under consideration.

A vector field $C=(C^1,\ldots,C^n)$ is said to be a \emph{conserved
vector} for the differential equations (\ref{F2}) if the equation
\begin{equation}
 \label{C1}
 D_i\big(C^i\big) = 0
\end{equation}
holds for any solution of Eq.~(\ref{F2}). 

If one of the independent variables is time, e.g., $x^n=t$,
then the conservation law is often written in the form
\begin{equation*}
 \ddt{E} = 0,
\end{equation*}
where
\begin{equation}
 \label{F9}
 E = \int\limits_{\R^{n-1}} C^n (x, u(x), u_{(1)}(x))\,\dx^1\cdots\dx^{n-1}.
\end{equation}
Accordingly, $C^n$ is termed the \emph{density} of the conservation law.

\subsection{Basic conservation theorem}

We will employ the recent general theorem \cite{Ibragimov:JMAA:2006b} on
a connection between symmetries and conservation laws for arbitrary
systems of $s$th-order partial differential equations
\begin{equation}
 F_\alpha \big(x, u, u_{(1)}, \ldots, u_{(s)}\big) = 0, \quad
 \alpha = 1, \ldots, m,
\end{equation}
where  $F_\alpha (x, u, u_{(1)},\ldots, u_{(s)})$ involves
$n$ independent variables $x = (x^1, \ldots, x^n)$
and $m$ dependent variables ${u=(u^1,\ldots,u^m)}$, $u = u(x)$
together with their derivatives up to an arbitrary order $s$.
For our purposes, we formulate the theorem in the case of systems of
first-order differential equations.
\begin{thm}
\label{bct.th}
(See Ref.~\onlinecite{Ibragimov:JMAA:2006b}, Theorem 3.5).  Let an
operator (\ref{F3}) be a symmetry of a system of first-order partial
differential equations
\begin{equation}
 \label{adeq1}
 F_\alpha \big(x, u, u_{(1)}\big) = 0, \quad
 \alpha = 1, \ldots, m.
\end{equation}
where $v=(v^1,\ldots,v^m)$. Then the quantities
\begin{equation}
 \label{F4}
 C^i = v^\beta \bigg[\xi^i F_\beta+ \big(\eta^\alpha - \xi^j
 u_j^\alpha\big) \frac{\partial F_\beta}{\partial
 u_i^\alpha}\bigg], \quad i = 1, \ldots, n,
\end{equation}
furnish a conserved vector $C=(C^1,\ldots,C^n)$ for the equations
(\ref{adeq1}) considered together with the adjoint system
\begin{equation}
\label{adeq2}
 F^*_\alpha \big(x, u, v,u_{(1)},v_{(1)}\big) \equiv
 \frac{\delta{\mathcal{L}}}{\delta u^\alpha}  = 0, \quad
 \alpha = 1, \ldots, m,
\end{equation}
where
\begin{equation*}
 \frac{\delta}{\delta u^\alpha} =
 \frac{\partial}{\partial u^\alpha} -
 D_i\,\frac{\partial}{\partial u^\alpha_i}
 \,,\quad \alpha=1, \ldots, m,
\end{equation*}
and $v = (v^1, \ldots, v^m)$ are are new dependent variables, i.e.,
${v=v(x)}$.
\end{thm}

\begin{rem}
\label{bct.re1}
The simultaneous system of equations (\ref{adeq1}) and (\ref{adeq2}) with
$2m$ dependent variables ${u=(u^1,\ldots,u^m)}$, ${v=(v^1,\ldots,v^m)}$ can be
obtained as the Euler-Lagrange equations (\ref{F2}) with the Lagrangian
\begin{equation}
\label{lagr}
 \mathcal{L} = v^\beta F_\beta \big(x, u, u_{(1)},\ldots, u_{(s)}\big)
\end{equation}

Indeed, 
\begin{align}
 &\frac{\delta\mathcal{L}}{\delta v^\alpha}
 = F_\alpha\big(x,u,u_{(1)}\big), \label{lagr1}\\
 &\frac{\delta\mathcal{L}}{\delta u^\alpha}
 = F^*_\alpha\big(x,u,v,u_{(1)},v_{(1)}\big).
\label{lagr2}
\end{align}
\end{rem}

\begin{rem}
\label{bct.re2}
The conserved quantities (\ref{F4}) can be written in terms of the
Lagrangian (\ref{lagr}) as follows:
\begin{equation}
\label{F4L}
 C^i = \mathcal{L} \xi^i + \big(\eta^\alpha - \xi^j
 u_j^\alpha\big) \frac{\partial \mathcal{L}}{\partial
 u_i^\alpha}\,.
\end{equation}
\end{rem}

\begin{rem}
\label{bct.re3}
If Eqs.~(\ref{adeq1}) have $r$ symmetries $X_1,\ldots,X_r$ of the form
(\ref{F3}),
\begin{align*}
 X_\mu  = \xi_\mu^i(x,u)\frac{\partial}{\partial x^i} + \eta_\mu^\alpha (x,u)
 \frac{\partial}{\partial u^\alpha}\,, \quad \mu = 1, \ldots, r,
\end{align*}
then Eqs.~(\ref{F4}) provide $r$ conserved vectors $C_1,\ldots,C_r$
with the components
\begin{align*}
 C_\mu^i = \mathcal{L}\xi_\mu^i + \big(\eta_\mu^\alpha - \xi_\mu^j
 u_j^\alpha\big) \frac{\partial\mathcal{L}}{\partial u_i^\alpha}\,,
 \quad \mu= 1, \ldots, r; \ \  i = 1, \ldots, n.
\end{align*}
\end{rem}

\section{Electromagnetic equations}

\subsection{Basic equations and the Lagrangian}

Adopting Dirac's ideas on the existence of magnetic monopoles
\cite{Dirac:PRSL:1931}, one can formulate a symmetrized version of Maxwell's
electromagnetic equations \cite{Schwinger:Science:1969}.  In SI units
and in microscopic (Lorentz) form, these equations are [cf.\
Ref.~\onlinecite[Eqs.~(1.50)]{Thide:2006}]:
\begin{subequations}
\label{eq:Maxwell-Dirac}
\begin{align}
 &\curl\vec{E} + \pddt{\vec{B}} + \mu_0\vec{j}_m=\vec{0},
   \label{IKT2.1}\\
 &\curl\vec{B} - \frac{1}{c^2}\pddt{\vec{E}} - \mu_0\vec{j}_e=\vec{0},
   \label{IKT2.2}\\
 &\div\vec{E} - \mu_0c^2\rho_e =0,\label{IKT2.3}\\
 &\div\vec{B} - \mu_0\rho_m =0, \label{IKT2.4}
\end{align}
\end{subequations}
together with the dual Ohm's law
\begin{equation}
\label{gohm}
\vec{j}_e = \sigma_e\vec{E}, \quad\vec{j}\,_m = \sigma_m\vec{B}.
\end{equation}
where $\sigma_m$ and $\sigma_e$ are constant scalar (rank zero)
quantities.  The first equation in (\ref{gohm}) is Ohm's law for
electric currents.  The second equation is a dual Ohm's law for magnetic
currents, that, for symmetry reasons, was introduced 
in Ref.~\onlinecite[Eqs.~(2.60)]{Thide:2006}; see also
Eq.~(5) in Ref.~\onlinecite{Meyer-Vernet:AJP:1982}, Eq.~(38) in
Ref.~\onlinecite{Olesen:PLB:1996}, and its generalization Eq.~(8) in
Ref.~\onlinecite{Coceal&al:EPL:1996}.

Now we substitute Eqs.~(\ref{gohm}) into
Eqs.~(\ref{IKT2.1})--(\ref{IKT2.2}).  The ensuing equations involve,
along with the light velocity $c$, three other constants, $\sigma_e,
\sigma_m$ and $\mu_0$. We eliminate two constants by setting
\begin{align}
 \label{dimless}
 &\tilde t = ct, \quad \widetilde{\vec{B}} = c \vec{B}, \quad
  \tilde{\sigma}_e = c \mu_0 \sigma_e, \quad
  \tilde{\sigma}_m = \frac{\mu_0}{c}\, \sigma_m, \notag\\[1ex]
 &\tilde{\rho}_e = c^2 \mu_0 \rho_e, \quad
  \tilde{\rho}_m = c \mu_0 \rho_m
\end{align}
and rewrite our basic Maxwell-Dirac equations (\ref{eq:Maxwell-Dirac}),
discarding tilde, as follows:
\begin{align}
\label{emeq}
\begin{split}
 &\curl\vec{E} + \pddt{\vec{B}} + \sigma_m\vec{B}=\vec0,\\
 &\curl\vec{B} - \pddt{\vec{E}} - \sigma_e\vec{E}=\vec0,\\
 &\div\vec{E} - \rho_e =0,\\
 &\div\vec{B} - \rho_m =0,
\end{split}
\end{align}
The system (\ref{emeq}) has eight equations for eight dependent
variables:  six coordinates of the electric and magnetic vector fields
$\vec{E}=(E^1,E^2,E^3)$ and $\vec{B}=(B^1,B^2,B^3)$, respectively, and
two scalar quantities, viz., the electric and magnetic monopole charge
densities $\rho_e$ and $\rho_m$.

Using the method of \citet{Ibragimov:JMAA:2006b} we write the Lagrangian
(\ref{lagr}) for Eqs.~(\ref{emeq}) in the following form:
\begin{align}
\label{emlag.1}
\begin{split}
 \mathcal{L} =&\vec{V}\cdot\Big(\curl\vec{E}+\pddt{\vec{B}}+\sigma_m\vec{B}\Big)
 + R_e \Big(\div\vec{E} - \rho_e\Big) \\
 &\mbox{}+\vec{W}\cdot\Big(\curl\vec{B} - \pddt{\vec{E}} - \sigma_e\vec{E}\Big)
 + R_m \Big(\div\vec{B} - \rho_m\Big),
\end{split}
\end{align}
where $\vec{V},\vec{W},R_{e},R_{m}$ are adjoint variables (we note in
passing that $\vec{V}$ is a pseudovector and $R_m$ a pseudoscalar).  With this
Lagrangian we have:
\begin{align}
\label{emlag.2}
\begin{split}
 & \frac{\delta\mathcal{L}}{\delta\vec{V}} = \curl\vec{E}
 + \pddt{\vec{B}} + \sigma_m\vec{B}, \quad
 \frac{\delta\mathcal{L}}{\delta R_e} = \div\vec{E} - \rho_e,\\
 &  \frac{\delta\mathcal{L}}{\delta\vec{W}} = \curl\vec{B}
  - \pddt{\vec{E}} - \sigma_e\vec{E},
  \quad  \frac{\delta\mathcal{L}}{\delta R_m} = \div\vec{B} - \rho_m,
\end{split}
\end{align}
and
\begin{align}
\label{emlag.3}
\begin{split}
 & \frac{\delta\mathcal{L}}{\delta \vec{E}} = \curl\vec{V} +
 \pddt{\vec{W}} - \sigma_e\vec{W} - \grad R_e,
 \quad  \frac{\delta\mathcal{L}}{\delta \rho_e} = - R_e, \\
 & \frac{\delta\mathcal{L}}{\delta \vec{B}} = \curl\vec{W} -
 \pddt{\vec{V}} + \sigma_m\vec{V} - \grad R_m,
 \quad  \frac{\delta\mathcal{L}}{\delta\rho_m} = - R_m.
\end{split}
\end{align}
It follows from Eqs.~(\ref{emlag.2})--(\ref{emlag.3}) that the
Euler-Lagrange equations (\ref{F2}) for the Lagrangian (\ref{emlag.1})
provide the electromagnetic equations (\ref{emeq}) and the following
adjoint equations for the new dependent variables
$\vec{V},\vec{W},R_e,R_m$:
\begin{align}
\label{emeq.adj}
\begin{split}
 &\curl\vec{V} + \pddt{\vec{W}} - \sigma_e\vec{W}=0,\\
 &\curl\vec{W} - \pddt{\vec{V}} + \sigma_m\vec{V}=0,\\
 &R_e = 0, \qquad R_m =0.
\end{split}
\end{align}

\begin{rem}
\label{emlag.4}
Let the spatial coordinates $x^1,x^2,x^3$ be $x,y,z$.  For computing the
variational derivatives $\delta\mathcal{L}/\delta\vec{E}$ and $\delta
\mathcal{L}/\delta\vec{B}$ in Eqs.~(\ref{emlag.3}), it is convenient to
use the coordinate representation of the Lagrangian (\ref{emlag.1}),
namely:
\begin{align}
\label{emlag.coord}
 \mathcal{L} & = V^1\,(E^3_y - E^2_z + B^1_t + \sigma_m B^1) +
 V^2\,(E^1_z - E^3_x + B^2_t + \sigma_m B^2)\notag\\[1ex]
 &+ V^3\,(E^2_x - E^1_y + B^3_t + \sigma_m B^3) + R_e (E^1_x + E^2_y + E^3_z - \rho_e)\\[1ex]
 &+ W^1\,(B^3_y - B^2_z - E^1_t - \sigma_e E^1)
  + W^2\,(B^1_z - B^3_x - E^2_t - \sigma_e E^2)\notag\\[1ex]
 &+ W^3\,(B^2_x - B^1_y - E^3_t - \sigma_e E^3) +
 R_m (B^1_x + B^2_y + B^3_z - \rho_m).\notag
\end{align}
\end{rem}

\subsection{Symmetries}

Eqs.~(\ref{emeq}) are invariant under the translations of time $t$ and
the position vector $\vec x = (x, y, z)$ as well as the simultaneous
rotations of the vectors $\vec{x},\vec{E}$ and $\vec{B}$ due to the
vector formulation of Eqs.~(\ref{emeq}).  These geometric
transformations provide the following seven infinitesimal symmetries:
\begin{align}
\label{emeq.sym}
\begin{split}
&\qquad X_0 = \pddt \,, \quad
 X_1 = \pddx \,, \quad
 X_2 = \pddy \,, \quad
 X_3 = \pddz \,,\\
 & X_{12} = y\frac{\partial}{\partial x} -
  x\frac{\partial}{\partial y} +
  E^2\frac{\partial}{\partial E^1} -
  E^1\frac{\partial}{\partial E^2} +
  B^2\frac{\partial}{\partial B^1} -
  B^1\frac{\partial}{\partial B^2}\,,\\
  & X_{13} = z\frac{\partial}{\partial x} -
  x\frac{\partial}{\partial z} +
  E^3\frac{\partial}{\partial E^1} -
  E^1\frac{\partial}{\partial E^3} +
  B^3\frac{\partial}{\partial B^1} -
  B^1\frac{\partial}{\partial B^3}\,,\\
 & X_{23} = z\frac{\partial}{\partial y} -
  y\frac{\partial}{\partial z} +
  E^3\frac{\partial}{\partial E^2} -
  E^2\frac{\partial}{\partial E^3} +
  B^3\frac{\partial}{\partial B^2} -
  B^2\frac{\partial}{\partial B^3}\,.
\end{split}
\end{align}
The infinitesimal symmetries for the adjoint system (\ref{emeq.adj}) are
obtained from (\ref{emeq.sym}) by replacing the vectors $\vec{E}$ and
$\vec{B}$ by $\vec{V}$ and $\vec{W}$, respectively.  Moreover, since
Eqs.~(\ref{emeq}) are homogeneous, they admit simultaneous
dilations of all dependent variables with the generator
\begin{equation}
\label{hom}
 T = \vec E \cdot \frac{\partial}{\partial \vec E}
 +\vec B \cdot \frac{\partial}{\partial \vec B} +
 \rho_e \frac{\partial}{\partial \rho_e} +
 \rho_m \frac{\partial}{\partial \rho_m}\,,
\end{equation}
where
\begin{equation*}
 \vec E \cdot \frac{\partial}{\partial \vec E} =
 \sum\limits_{i=1}^3 E^i \frac{\partial}{\partial E^i}\,,
 \quad
 \vec B \cdot \frac{\partial}{\partial \vec B}
 = \sum\limits_{i=1}^3 B^i \frac{\partial}{\partial B^i}\,.
\end{equation*}

Recall that the Maxwell equations in vacuum admit also the one-parameter
group of \emph{Heaviside-Larmor-Rainich duality transformations}
\begin{equation}
\label{eq:HLR}
 \overline{\vec E}= \vec E \cos\alpha-\vec B\sin\alpha, \quad
 \overline{\vec B}=\vec E\sin\alpha+\vec B\cos\alpha
\end{equation}
with the generator
\begin{equation*}
 X = \vec E \cdot \frac{\partial}{\partial \vec B}
  -\vec B \cdot \frac{\partial}{\partial \vec E} \equiv
 \sum\limits_{i=1}^3 \bigg(E^i \frac{\partial}{\partial B^i}
 - B^i \frac{\partial}{\partial E^i}\bigg)\,.
\end{equation*}
Also recall that the ``mixing angle'' $\alpha$ in (\ref{eq:HLR}) is
a pseudoscalar.

It was shown in \cite{Ibragimov:JMAA:2006b} that the group
(\ref{eq:HLR}) provides the conservation of energy for the Maxwell
equations.  Let us clarify whether Eqs.~(\ref{emeq}) admit a similar
group.  Let therefore
\begin{equation}
\label{IKT2.7}
  X = \vec E \cdot \frac{\partial}{\partial \vec B}
  -\vec B \cdot \frac{\partial}{\partial \vec E} +
 \rho_e \frac{\partial}{\partial \rho_m} -
 \rho_m \frac{\partial}{\partial \rho_e}\,.
 \end{equation}
The prolongation  of the operator (\ref{IKT2.7}) is written
\begin{multline}
\label{IKT2.8}
 X = \vec E\cdot\frac{\partial}{\partial\vec B}
 -\vec B\cdot\frac{\partial}{\partial \vec E}
 +\rho_e\frac{\partial}{\partial\rho_m}
 -\rho_m\frac{\partial}{\partial\rho_e} \\
 +\vec E_t\cdot\frac{\partial}{\partial\vec B_t} 
 -\vec B_t\cdot\frac{\partial}{\partial\vec E_t}
 +\vec E_x\cdot\frac{\partial}{\partial\vec B_x}
 -\vec B_x\cdot\frac{\partial}{\partial\vec E_x} \\
 +\vec E_y\cdot\frac{\partial}{\partial\vec B_y} 
 -\vec B_y\cdot\frac{\partial}{\partial\vec E_y}
 +\vec E_z\cdot\frac{\partial}{\partial\vec B_z}
 -\vec B_z\cdot\frac{\partial}{\partial\vec E_z}\,.
\end{multline}
Reckoning shows that the operator (\ref{IKT2.8}) acts on the left-hand
sides of Eqs.~(\ref{emeq}) as follows:
\begin{gather*}
 X\big(\curl\vec{E} + \vec{B}_t +
  \sigma_m \vec{B}\big) = - \big(\curl\vec{B} -
 \vec{E}_t - \sigma_m  \vec{E}\big),\\
 X\big(\curl\vec{B} - \vec{E}_t
 - \sigma_e \vec{E}\big) = \curl\vec{E} + \vec{B}_t + \sigma_e \vec{B},\\
 X\big(\div\vec{E} - \rho_e\big) =
 - \big(\div\vec{B} - \rho_m\big),\\
 X\big(\div\vec{B} - \rho_m\big) = \div\vec{E} - \rho_e.
\end{gather*}
It follows that the operator (\ref{IKT2.7}) is admitted by Eqs.
 (\ref{emeq}) only in the case
\begin{equation}
\label{emlag.5}
 \sigma_m = \sigma_e.
\end{equation}

\section{Conservation laws}

\subsection{Derivation of conservation laws}
\label{dcl}

We will write the conservation law~(\ref{C1}) in the form
\begin{equation}
\label{IKT2.9}
 D_t(\tau) + \mathrm{div}\,\bm{\chi} = 0,
\end{equation}
where the pseudoscalar $\tau$ is the density of the conservation
law~(\ref{IKT2.9}), the pseudovector current
${\bm{\chi}=(\chi^1,\chi^2,\chi^3)}$, and
\begin{equation*}
 \mathrm{div}\,\bm{\chi} \equiv \div\bm{\chi} =
 D_x(\chi^1) + D_y(\chi^2) + D_z(\chi^3).
\end{equation*}

Let us find the conservation law furnished by the symmetry
(\ref{IKT2.7}) when the condition (\ref{emlag.5}) is satisfied,
$\sigma_m=\sigma_e$.  Applying the formula (\ref{F4}) to the symmetry
(\ref{IKT2.7}) and to the Lagrangian (\ref{emlag.1}), we obtain the
following density of the conservation law~(\ref{IKT2.9}):
\begin{equation*}
 \tau = \vec{E}\cdot\frac{\partial\mathcal{L}}{\partial \vec B_t}
 -\vec{B}\cdot\frac{\partial\mathcal{L}}{\partial \vec E_t}
 =\vec{E}\cdot\vec{V} + \vec{B}\cdot\vec{W}.
\end{equation*}
Thus,
\begin{equation}
\label{IKT2.10}
 \tau = \vec{E}\cdot\vec{V} + \vec{B}\cdot\vec{W}.
\end{equation}
The pseudovector $\bm{\chi}$ is obtained likewise. For example, using
the Lagrangian in the form (\ref{emlag.coord}), we have:
\begin{equation*}
 \chi^1 = \vec{E}\cdot\frac{\partial\mathcal{L}}{\partial \vec B_x}
 -\vec{B}\cdot\frac{\partial\mathcal{L}}{\partial\vec E_x}
 = E^2 W^3 - E^3 W^2 - B^2 V^3 + B^3 V^2.
\end{equation*}
The other coordinates of $\bm{\chi}$ are computed likewise, and the final
result is:
\begin{equation}
\label{IKT2.11}
 \bm{\chi} = (\vec{E}\cross\vec{W}) - (\vec{B}\cross\vec{V}).
\end{equation}
One can readily verify that (\ref{IKT2.10}) and (\ref{IKT2.11}) provide
a conservation law for Eqs.~(\ref{emeq}) considered together with the
adjoint equations (\ref{emeq.adj}). Indeed, using the well-known
formula $\div(\vec{a}\cross\vec{b})=\vec{b}\cdot(\curl\vec{a})
-\vec{a}\cdot(\curl\vec{b})$ and Eqs.~(\ref{emeq}) and (\ref{emeq.adj}),
we obtain:
\begin{align*}
 D_t (\tau) &= \vec{E}_t\cdot\vec{V} + \vec{E}\cdot\vec{V}_t
 + \vec{B}_t\cdot\vec{W}+ \vec{B}\cdot\vec{W}_t\\
 &= \vec{V}\cdot(\curl\vec{B} - \sigma_e\vec{E})
 + \vec{E}\cdot(\curl\vec{W} + \sigma_m\vec{V})\\
 &\quad\mbox{}- \vec{W}\cdot(\curl\vec{E} + \sigma_m\vec{B})
 - \vec{B}\cdot(\curl\vec{V} - \sigma_e
 \vec{W}),\notag\\[1.5ex]
 \div\bm{\chi} &= \div(\vec{E}\cross\vec{W}) -
 \div\vec{B}\cross\vec{V})\\
 &= \vec{W}\cdot(\curl\vec{E})
  - \vec{E}\cdot(\curl\vec{W}) \\
 &\quad\mbox{}- \vec{V}\cdot(\curl\vec{B})
  + \vec{B}\cdot(\curl\vec{V}).
\end{align*}
Whence,
\begin{equation*}
 D_t (\tau) + \mathrm{div}\, \bm{\chi} = (\sigma_m
 - \sigma_e) (\vec{E}\cdot\vec{V} - \vec{B}\cdot\vec{W}).
\end{equation*}
It follows again that the conservation law is valid only if
${\sigma_m=\sigma_e}$.

\begin{rem}
\label{emr.1}
The conservation law given by (\ref{IKT2.10})--(\ref{IKT2.11}) depends
on solutions $(\vec V,\vec W)$ of the adjoint system (\ref{emeq.adj}).
However, substituting into Eqs.~(\ref{IKT2.10}) and (\ref{IKT2.11}) any
particular solution $(\vec V,\vec W)$ of the adjoint system
(\ref{emeq.adj}) with $\sigma_m=\sigma_e$, one obtains the conservation
law for Eqs.~(\ref{emeq}) not involving $\vec V$ and $\vec W$. Let us
denote $\sigma_m=\sigma_e=\sigma$ and take, e.g., the following
simple solution of the adjoint system (\ref{emeq.adj}):
\begin{equation*}
 V^1 = e^{\sigma t}, \ V^2 = V^3 = 0; \quad
 W^1 = e^{\sigma t}, \ W^2 = W^3 = 0.
\end{equation*}
Then Eqs.~(\ref{IKT2.10})--(\ref{IKT2.11}) yield:
\begin{gather*}
 \tau = (E^1 + B^1)e^{\sigma t} \\
 \chi^1 = 0 \quad
 \chi^2 = (E^3 - B^3)e^{\sigma t} \quad
 \chi^3 = (B^2 - E^2)e^{\sigma t}\,.
\end{gather*}
\end{rem}

\begin{rem}
The operator (\ref{IKT2.7}) generates the one-parameter group
\begin{gather*}
 \overline{\vec E}= \vec E \cos\alpha-\vec B\sin\alpha, \quad
 \overline{\vec B}=\vec E\sin\alpha+\vec B\cos\alpha,\\
 \overline{\rho}\,^e= \rho_e \cos\alpha-\rho_m\sin\alpha, \quad
 \overline{\rho}\,^m= \rho_e\sin\alpha+\rho_m\cos\alpha\,.
\end{gather*}
where, again, the ``mixing angle'' $\alpha$ is a pseudoscalar.
\end{rem}

\begin{rem}
In the original variables used in Eqs.~(\ref{IKT2.1})--(\ref{IKT2.2})
and (\ref{gohm}), the operator (\ref{IKT2.7}) is written:
\begin{equation*}
 X = \frac{1}{c}\,\vec E\cdot\frac{\partial}{\partial\vec B}
 -c \,\vec B\cdot\frac{\partial}{\partial\vec E} +
 c\,\rho_e \frac{\partial}{\partial\rho_m} -
 \frac{1}{c}\,\rho_m \frac{\partial}{\partial\rho_e}\,.
\end{equation*}
\end{rem}

Applying similar calculations to the generator (\ref{hom}) of
the dilation group provides the conservation law with
\begin{equation}
\label{hom.1}
 \tau = \vec{B}\cdot\vec{V} - \vec{E}\cdot\vec{W}, \quad
 \bm{\chi} = (\vec{E}\cross\vec{V}) + (\vec{B}\cross\vec{W}).
\end{equation}
This conservation law is valid for arbitrary $\sigma_m$ and
$\sigma_e$. Indeed,
\begin{align*}
  D_t (\tau) &= \vec{B}_t\cdot \vec{V} + \vec{B}\cdot\vec{V}_t -
  \vec{E}_t\cdot \vec{W} - \vec{E}\cdot\vec{W}_t\\
 &= -\vec{V}\cdot(\curl\vec{E} + \sigma_m \vec{B})
 + \vec{B}\cdot(\curl\vec{W} + \sigma_m\vec{V})\\
 &\quad\mbox{}-\vec{W}\cdot(\curl\vec{B} - \sigma_e\vec{E})
 + \vec{E}\cdot(\curl\vec{V} - \sigma_e\vec{W})\\
 &= -\vec{V}\cdot(\curl\vec{E})
 + \vec{B}\cdot(\curl\vec{W})\\
 &\quad\mbox{}-\vec{W}\cdot(\curl\vec{B}) 
  + \vec{E}\cdot(\curl\vec{V})\,,\\[1.5ex]
 \div\bm{\chi} &= \div(\vec{E}\cross\vec{V}) + \div(\vec{B}\cross\vec{W})\\
 &= \vec{V}\cdot(\curl\vec{E}) -\vec{E}\cdot(\curl\vec{V})\\
 &\quad\mbox{}+\vec{W}\cdot(\curl\vec{B}) - \vec{B}\cdot(\curl\vec{W})\,.
\end{align*}
Hence, $D_t(\tau)+\mathrm{div}\,\bm{\chi}=0$.

Let us find the conservation law provided by the symmetry
$X_0={\partial}/{\partial t}$ from (\ref{emeq.sym}).  Formula (\ref{F4})
yields:
\begin{align*}
 \tau = \mathcal{L} - \vec E_t\cdot\frac{\partial\mathcal{L}}{\partial\vec E_t}
 -\vec B_t\cdot \frac{\partial\mathcal{L}}{\partial \vec B_t}
 = \mathcal{L} + \vec E_t \cdot \vec{W} - \vec B_t \cdot \vec{V}.
\end{align*}
Since the Lagrangian $\mathcal{L}$ given by (\ref{emlag.1}) vanishes on
the solutions of Eqs.~(\ref{emeq}), we can take
\begin{equation}
 \label{IKT2.12A}
 \tau = \vec E_t\cdot\vec{W} - \vec B_t\cdot\vec{V}
\end{equation}
or
\begin{equation}
\label{IKT2.12}
 \tau = \vec{W}\cdot[(\curl\bm{B}) - \sigma_e\vec{E}]
  + \vec{V}\cdot[(\curl\vec{E}) + \sigma_m \vec{B}].
 \end{equation}
Let us calculate the pseudovector $\bm{\chi}$. Formula (\ref{F4}) yields:
\begin{equation*}
 \chi^1 = - \vec{E}_t\cdot\frac{\partial\mathcal{L}}{\partial \vec E_x}
 -\vec{B}_t\cdot \frac{\partial\mathcal{L}}{\partial \vec B_x}\,.
\end{equation*}
Using the Lagrangian in the form (\ref{emlag.coord}), we have:
\begin{equation*}
 \chi^1 = - E^2_t V^3 + E^3_t V^2 - B^2_t W^3 + B^3_t W^2.
\end{equation*}
The other coordinates of $\bm{\chi}$ are computed similarly, and
the final result is
\begin{equation}
\label{IKT2.13}
 \bm{\chi} = (\vec{V}\cross\vec{E}_t) + (\vec{W}\cross\vec{B}_t).
\end{equation}
Thus, the time translational invariance of Eqs.~(\ref{emeq}) leads to
the conservation law (\ref{IKT2.9}) with $\tau$ and $\bm{\chi}$ given
by (\ref{IKT2.12}) and (\ref{IKT2.13}), respectively.

\begin{rem}
Let us substitute in Eqs.~(\ref{IKT2.12}) and (\ref{IKT2.13})
the following simple solution of the adjoint
system (cf.~Remark \ref{emr.1}):
\begin{equation*}
 V^1 = e^{\sigma_m t}, \ V^2 = V^3 = 0; \quad
 W^1 = e^{\sigma_e t}, \ W^2 = W^3 = 0.
\end{equation*}
 Then Eqs.~(\ref{IKT2.12})--(\ref{IKT2.13}) yield:
\begin{gather*}
 \tau = (B^3_y - B^2_z - \sigma_e E^1) e^{\sigma_e t}
 + (E^3_y - E^2_z + \sigma_m B^1) e^{\sigma_m t},\\
  \chi^1 = 0, \quad \chi^2 = - E^3_t e^{\sigma_m t}
  - B^3_t e^{\sigma_e t}, \quad
  \chi^3 = E^2_t e^{\sigma_m t} + B^2_t e^{\sigma_e t}.
\end{gather*}
\end{rem}
 %%%%%%%%%%%%%%%%%%%%%%%%%%%%%%%%%%%%%%%%%%%%%%%%%%%%%%%%%%

The conservation law provided by the symmetry $X_1=\partial/\partial x$
from (\ref{emeq.sym}) has the following density:
\begin{equation*}
 \tau = - \vec E_x\cdot \frac{\partial\mathcal{L}}{\partial \vec E_t}
 -\vec B_x\cdot \frac{\partial\mathcal{L}}{\partial \vec B_t}
 =  \vec E_x\cdot \vec{W} - \vec B_x \cdot \vec{V}.
\end{equation*}
 %Invoking the expression (\ref{emlag.1}) for the Lagrangian
 %${\mathcal{L}$ we have:
 %\begin{equation}
 %\label{IKT2.121}
 %\tau = \vec{W}  \cdot [(\nabla \times \vec{B}) - \sigma_e \vec{E}]
 % + \vec{V}  \cdot [(\nabla \times \vec{E}) + \sigma_m \vec{B}].
% \end{equation}
For the pseudovector $\bm{\chi}$ the formula (\ref{F4}) yields:
\begin{equation*}
\chi^1 = \mathcal{L}- \vec{E}_x\cdot \frac{\partial\mathcal{L}}{\partial
\vec E_x}
 -\vec{B}_x\cdot \frac{\partial\mathcal{L}}{\partial\vec B_x}\,.
\end{equation*}
Using the Lagrangian in the form (\ref{emlag.coord}), we have:
\begin{equation*}
 \chi^1
 = \mathcal{L}- E^2_x V^3 + E^3_x V^2 - B^2_x W^3 + B^3_x W^2.
\end{equation*}
The other coordinates of $\bm{\chi}$ are calculated similarly:
\begin{gather*}
 \chi^2
 =  E^1_x V^3 - E^3_x V^1 + B^1_x W^3 - B^3_x W^1,\\
 \chi^3
 =  -E^1_x V^2 + E^2_x V^1 - B^1_x W^2 + B^2_x W^1.
\end{gather*}
We can ignore $\mathcal{L}$ in $\chi^1$ since $D_x\mathcal{L}=0$ on
solutions of Eqs.~(\ref{emeq}) and (\ref{emeq.adj}), and
the final result is
\begin{equation}
\label{IKT2.131}
 \bm{\chi} = (\vec{V}\cross\vec{E}_x) + (\vec{W}\cross\vec{B}_x).
\end{equation}
Replacing $x$ by $y$ and $z$ we obtain the following conservation laws
corresponding to $X_2=\partial/\partial y$ and $X_3=\partial/\partial z$,
respectively:
\begin{equation*}
 \tau =\vec E_y\cdot\vec{W} - \vec B_y\cdot\vec{V},\quad
\bm{\chi} = (\vec{V}\cross\vec{E}_y) + (\vec{W}\cross\vec{B}_y)
\end{equation*}
and
\begin{equation*}
 \tau =\vec E_z\cdot\vec{W} - \vec B_z\cdot\vec{V},\quad
\bm{\chi} = (\vec{V}\cross\vec{E}_z) + (\vec{W}\cross\vec{B}_z).
\end{equation*}

Applying Formula (\ref{F4}) to the symmetry $X_{12}$ and to the
Lagrangian (\ref{emlag.coord}), we obtain the following density of the
conservation law:
\begin{align*}
 \tau =& E^{2}\frac{\partial\mathcal{L}}{\partial  E^{1}_t}-
 E^{1}\frac{\partial\mathcal{L}}{\partial E^{2}_t}
 + (x\vec E_y -y \vec E_x)\cdot \frac{\partial\mathcal{L}}{\partial\vec E_t} \\
 &\mbox{}+ B^{2}\frac{\partial\mathcal{L}}{\partial  B^{1}_t}-
 B^{1}\frac{\partial\mathcal{L}}{\partial \vec B^{2}_t}
 + (x\vec B_y -y \vec B_x)\cdot \frac{\partial\mathcal{L}}{\partial\vec B_t}\\
 =& W^{2}E^{1}-W^{1}E^{2}
  + (y\vec E_x -x\vec E_y)\cdot\vec{W} \\
 &\mbox{} - (V^{2}B^{1}-V^{1}B^{2})- (y\vec B_x -x\vec B_y) \cdot \vec{V}.
\end{align*}
The densities of the conservation laws for $X_{13}$ and $X_{23}$
are
\begin{multline*}
 \tau = W^{3}E^{1}-W^{1}E^{3}
  + (z\vec E_x -x\vec E_z)\cdot \vec{W} \\
  - (V^{3}B^{1}-V^{1}B^{3})- (z\vec B_x -x\vec B_z) \cdot \vec{V}
\end{multline*}
and
\begin{multline*}
 \tau
 = W^{3}E^{2}-W^{2}E^{3}
  +  (z\vec E_y -y\vec E_z)\cdot \vec{W} \\
 \mbox{} - (V^{3}B^{2}-V^{2}B^{3})- (z\vec B_y -y\vec B_z) \cdot
  \vec{V},
\end{multline*}
respectively.  Finally, the densities of conservation laws corresponding
to the rotation generators $X_{ij}$ can be written as one vector:
\begin{equation}
\label{IKT2.132}
 \vec\tau =\vec W\cross\vec{E}
  + \vec{W}\cdot(\vec x\cross\grad)\vec{E}  -
  \vec V\cross\vec{B}- \vec{V}\cdot(\vec x\cross\grad)\vec{B},
\end{equation}
where $\vec x=(x,y,z)$.

The operator $X_{12}$ provides the following pseudovector $\bm{\chi}$:
\begin{align*}
\chi^1 =&-V^{3}E^{1}- y(E^2_x V^3 - E^3_x V^2)+x(E^2_y V^3 - E^3_y V^2) \\
 &\mbox{}-W^{3}B^{1} - y(B^2_x W^3 - B^3_x W^2)+x(B^2_y W^3 - B^3_yW^2), \\
\chi^2 =&-V^{3}E^{2}+ y(E^1_x V^3 - E^3_x V^1)-x(E^1_y V^3 - E^3_y V^1) \\
 &\mbox{}-W^{3}B^{2} + y(B^1_x W^3 - B^3_x W^1)-x(B^1_y W^3 - B^3_yW^1),\\
\chi^3 =& V^{1}E^{1}+ V^{2}E^{2}- y(E^1_x V^2 - E^2_x V^1)+x(E^1_y V^2 - E^2_y V^1) \\
 &\mbox{}+W^{1}B^{1}+ W^{2}B^{2}- y(B^1_x W^2 - B^2_x W^1)+x(B^1_y W^2 - B^2_y W^1)\,.
\end{align*}
The pseudovector $\bm{\chi}$ for the operator $X_{13}$ has the
following form:
\begin{align*}
 \chi^1 =& V^{2}E^{1}- z(E^2_x V^3 - E^3_x V^2)+x(E^2_z V^3 - E^3_z V^2) \\
 &\mbox{}+W^{2}B^{1} - z(B^2_x W^3 - B^3_x W^2)+x(B^2_z W^3 - B^3_zW^2),\\
 \chi^2 =&-V^{1}E^{1}-V^{3}E^{3}+ z(E^1_x V^3 - E^3_x V^1)-x(E^1_z V^3 - E^3_z V^1) \\
 &\mbox{}-W^{1}B^{1}-W^{3}B^{3} + z(B^1_x W^3 - B^3_x W^1)-x(B^1_z W^3 - B^3_zW^1),\\
 \chi^3 =&V^{2}E^{3}- z(E^1_x V^2 - E^2_x V^1)+x(E^1_z V^2 - E^2_z V^1) \\
 &\mbox{} + W^{2}B^{3}- z(B^1_x W^2 - B^2_x W^1)+x(B^1_z W^2 - B^2_z W^1)\,.
\end{align*}
The operator $X_{23}$ provides the following pseudovector $\bm{\chi}$:
\begin{align*}
 \chi^1 =& V^{2}E^{2}+V^{3}E^{3}- z(E^2_y V^3 - E^3_y V^2)+y(E^2_z V^3 - E^3_z V^2) \\
 &\mbox{}+W^{2}B^{2}+W^{3}B^{3}- z(B^2_y W^3 - B^3_y W^2)+y(B^2_z W^3 - B^3_zW^2),\\
 \chi^2 =&-V^{1}E^{2}+ z(E^1_y V^3 - E^3_y V^1)-y(E^1_z V^3 - E^3_z V^1) \\
 &\mbox{} -W^{1}B^{2} + z(B^1_y W^3 - B^3_y W^1)-y(B^1_z W^3 - B^3_zW^1)\,,\\
 \chi^3 =&-V^{1}E^{3}- z(E^1_y V^2 - E^2_y V^1)+y(E^1_z V^2 - E^2_z V^1) \\
 &\mbox{} -W^{1}B^{3}- z(B^1_yW^2 - B^2_y W^1)+y(B^1_z W^2 - B^2_z W^1)\,.
\end{align*}

\subsection{Two-solution representation of conservation laws}
 \label{tsr}

The conserved quantities obtained in Subsection \ref{dcl} involve solutions
$\vec{V}, \vec{W}$ of the adjoint equations (\ref{emeq.adj}).  It may be
useful for applications to give an alternative representation of the
conserved quantities in terms of the electric and magnetic vector fields
$\vec{E}, \vec{B}$ only.

We suggest here one possibility based on the observation that one
can satisfy the adjoint system (\ref{emeq.adj}) by letting
\begin{align}
\label{tsr1}
\begin{split}
 & \vec{V}(\vec{x},t) =  \vec{B}(\vec{x},-t), \\
 & \vec{W}(\vec{x},t) =  \vec{E}(\vec{x},-t), \\
 & R_e(\bm{x},t) = \div\vec{E}(\vec{x},-t) -\rho_e(\vec{x},-t), \\
 & R_m(\bm{x},t) = \div\vec{B}(\vec{x},-t) -\rho_m(\vec{x},-t),
\end{split}
\end{align}
where $\vec{E}(\vec{x},s),\vec{B}(\vec{x},s)$ solve Eqs.~(\ref{emeq})
with $s=-t$.  Indeed, employing the substitution (\ref{tsr1}) and the
notation $s=-t$ we have
\begin{align}
%\label{emeq.adjA}
\begin{split}
 &\curl\vec{V} + \pddt{\vec{W}}- \sigma_e\vec{W} \\
 &\qquad= \curl\vec{B}(\vec{x},s) + \pdds{\vec{E}(\vec{x},s)}\pddt{s} 
  \mbox{}- \sigma_e\,\vec{E}(\vec{x},s)\,, \\
 &\curl\vec{W} - \pddt{\vec{V}} + \sigma_m \vec{V} \\
 &\qquad = \curl\vec{E}(\vec{x},s) - \pdds{\vec{B}(\vec{x},s)} \pddt{s} 
  + \sigma_m\vec{B}(\vec{x},s)\,, \\
 & R_e = \div\vec{E}(\vec{x},s) - \rho_e(\vec{x},s)=0\,, \\
 & R_m =\div\vec{B}(\vec{x},s) - \rho_m(\vec{x},s)\,.
\end{split}
\end{align}
Hence, the adjoint equations (\ref{emeq.adj}) reduce to (\ref{emeq}):
\begin{align}
\label{tsr2}
\begin{split}
 & \curl\vec{E}(\vec{x},s) + \pdds{\vec{B}(\vec{x},s)}
  + \sigma_m\vec{B}(\vec{x},s)=0, \\
 & \div\vec{B}(\vec{x},s) - \pdds{\vec{E}(\vec{x},s)} -
   \sigma_e\vec{E}(\vec{x},s)=0, \\
 & \div\vec{E}(\vec{x},s) - \rho_e(\vec{x},s)=0, \\
 & \div\vec{B}(\vec{x},s) - \rho_m(\vec{x},s)=0\,.
\end{split}
\end{align}

Let $\big(\vec{E}(\vec{x},t), \vec{B}(\vec{x},t)\big)$ and
$\big(\vec{E}'(\vec{x},t),\vec{B}'(\vec{x},t)\big)$ be any two solutions
of the electromagnetic equations (\ref{emeq}).  Substituting in
(\ref{tsr1}) the solution $\big(\vec{E}',\vec{B}'\big),$ we obtain the
\emph{two-solution representations} of the conservation laws.  For
example, the conservation law given by (\ref{IKT2.10})--(\ref{IKT2.11})
has in this representation the following coordinates:
\begin{align}
\label{IKT2.10T}
\begin{split}
 & \tau = \vec{E}(\vec{x},t)\cdot\vec{B}'(\vec{x},-t) +
 \vec{B}(\vec{x},t)\cdot\vec{E}'(\vec{x},-t),\\
& \bm{\chi} = [\vec{E}(\vec{x},t)\cross\vec{E}'(\vec{x},-t)] -
 [\vec{B}(\vec{x},t)\cross\vec{B}'(\vec{x},-t)]\,.
\end{split}
\end{align}
In particular, if the solutions
$\big(\vec{E}(\vec{x},t),\vec{B},t)\big)$ are identical, (\ref{IKT2.10T})
provides the \emph{one-solution representation}:
\begin{align}
\label{IKT2.10Ta}
\begin{split}
 &\tau = \vec{E}(\vec{x},t)\cdot\vec{B}(\vec{x},-t) +
 \vec{B}(\vec{x}, t)\cdot\vec{E}(\vec{x},-t),\\
& \bm{\chi} = [\vec{E}(\vec{x},t)\cross\vec{E}(\vec{x},-t)] -
 [\vec{B}(\vec{x},t)\cross\vec{B}(\vec{x},-t)]\,.
\end{split}
\end{align}

All other conservation laws can be treated likewise, e.g., the
conservation law given by (\ref{IKT2.12A}) and (\ref{IKT2.13}) has the
following two-solution representation:
\begin{align}
\label{IKT2.12AT}
\begin{split}
 & \tau = \vec E_t(\vec{x},t)\cdot\vec{E}'(\vec{x},-t)
  -\vec B_t(\vec{x},t)\cdot\vec{B}'(\vec{x},-t),\\
 & \bm{\chi} = [\vec{B}'(\vec{x},-t)\cross\vec{E}_t(\vec{x},t)]
 + [\vec{E}'(\vec{x},-t)\cross\vec{B}_t(\vec{x},t)]\,.
\end{split}
\end{align}

\begin{acknowledgments}
One of the authors (B.\,T.) gratefully acknowledges the financial
support from the Swedish Governmental Agency for Innovation Systems
(VINNOVA).  
\end{acknowledgments}

%%%%%%%%%%%%%%%%%%%%%%%%%%%%%%%%%%%%%%%%%%%%%%%%%%%%%%%%%%%%%%%%%%%%%%

%\bibliographystyle{amsplain}
\bibliographystyle{apsrev}
\bibliography{nail}

\end{document}